\documentclass{JHEP3} 

\usepackage{amsmath,latexsym}
\usepackage[dvips]{graphicx}
\usepackage{longtable}
\usepackage{supertabular}
\usepackage{color}

\providecommand{\openone}{\leavevmode\hbox{\small1\kern-3.5pt\normalsize1}}

\newcommand\fverb{\setbox\pippobox=\hbox\bgroup\verb}
\newcommand\fverbdo{\egroup\medskip\noindent%
                        \fbox{\unhbox\pippobox}\ }
\newcommand\fverbit{\egroup\item[\fbox{\unhbox\pippobox}]}
\newbox\pippobox

\preprint{FTUV/08-0219, IFIC/08-10, MPP-2008-13, UAB-FT-638}

\title{OPE--R\boldmath{$\chi$}T matching at order \boldmath{$\alpha_s$}:
 hard gluonic corrections to three--point Green functions}

\author{Matthias Jamin\\
        Instituci\'o Catalana de Recerca i Estudis Avan\c{c}ats (ICREA), IFAE,\\
        Theoretical Physics Group, UA Barcelona,
        E-08193 Bellaterra, Barcelona, Spain\\
        E-mail: \email{jamin@ifae.es}}

\author{Vicent Mateu\thanks{On leave from: Departament~de~F\'\i sica~Te\`orica,
        IFIC, Universitat~de~Val\`encia-CSIC, Apt.~Correus 22085,
        E-46071 Val\`encia, Spain.}\\
        Max-Planck-Institut f\"ur Physik (Werner-Heisenberg-Institut),\\
        F\"ohringer Ring 6, D-80805 M\"unchen, Germany\\
        E-mail: \email{mateu@mppmu.mpg.de}}

\abstract{In this work we push the matching between the QCD operator product
expansion (OPE) and resonance chiral theory (R$\chi$T) to order $\alpha_{s}$.
To this end we compute two-- and three--point QCD Green functions (GFs) in both
theories and compare the results. GFs which are order parameters of chiral
symmetry breaking make this matching more transparent and thus we concentrate
on those. On the OPE side one needs to calculate the hard--gluon virtual
corrections to the quark condensate, and in particular for three--point GFs
this computation was hitherto missing. We also discuss the need for including
the infinite tower of hadronic states in the hadronic representation of the GF
when non--analytic terms such as logarithms are present in the OPE Wilson
coefficients.}


\begin{document}

\section{Introduction\label{sec:Introduction}}

Green functions (GFs for short) of colour singlet local QCD currents
have proved to be an essential tool for a successful description of hadronic
physics. Confinement binds quarks and gluons into colour singlet hadrons
and hence it makes little sense to compute matrix elements with those
fundamental particles as asymptotic states. The process of hadronisation
takes place at an energy scale of the order of $\Lambda_{\mathrm{QCD}}$
and thus it is essentially non--perturbative. QCD currents can be written as
\begin{equation}
J_{\Gamma}^{a}(x) \,\equiv\; \,:\!\bar{q}_{\alpha iA}(x)\Gamma_{\alpha\beta}
\biggl(\frac{\lambda^{a}}{2}\biggr)_{ij}\!q_{\beta jA}(x)\!:\, \;=\;
\,:\!\bar{q}(x)\,\Gamma\,\frac{\lambda^{a}}{2}\, q(x)\!: \,,
\end{equation}
where $\Gamma$ and $\lambda^{a}$ are spin and flavour matrices, respectively,
and $A$ denotes a (summed) colour index. The $\Gamma$ matrix stands for $2\, I$,
$2\,i\,\gamma_{5}$, $\gamma^{\mu}$, $\gamma^{\mu}\,\gamma_{5}$ and
$\sigma^{\mu\nu}$ in the case of scalar, pseudoscalar, vector, axial--vector
and tensor sources, respectively.\footnote{In our conventions
$\gamma_{5}=i\,\gamma_{0}\gamma_{1}\gamma_{2}\gamma_{3}$ and
$\sigma^{\mu\nu}=\frac{i}{2}\left[\gamma^{\mu},\gamma^{\nu}\right]$.}
The currents $J_{\Gamma}^{a}(x)$ qualify as interpolating fields for mesonic
states $H_{\Gamma}^{a}$ in the sense of having non--vanishing matrix elements
between the vacuum and the hadronic state:
$\langle\, H_{J}^{a}\,|\,J_{\Gamma}^{b}\,|\,0\,\rangle\neq0$. This permits to
obtain matrix elements of hadronic states from GFs through the LSZ reduction
formula. Let us define the two-- and three--point GFs in momentum space:
\begin{eqnarray}
\Pi_{12}^{ab}(p) &\,\equiv\,& i\!\int\mathrm{d}^{4}x\, e^{ip\cdot x}\langle\,0
\,|\,T\{ J^a_{1}(x)J^b_{2}(0)\}|\,0\,\rangle\,,\nonumber \\
\Pi_{123}^{abc}(p,q) &\,\equiv\,& i^{2}\!\int\mathrm{d}^{4}x\,\mathrm{d}^{4}y\,
e^{i(p\cdot x+q\cdot y)}\langle\,0\,|\,T\{ J^a_{1}(x)J^b_{2}(y)\, J^c_{3}(0)\}|
\,0\,\rangle\,.
\end{eqnarray}
These definitions can be trivially generalised for more currents.

QCD exhibits a spontaneous breakdown of the chiral symmetry due to quantum
effects.  It is believed that the operator responsible for this phenomenon is
the quark condensate $\langle\bar{q}q\rangle$, which acquires a non--vanishing
expectation value. Phenomenologically, this feature is reflected through the
appearance in the spectrum of the so called pseudo--Goldstone bosons. At very
low energies (long distances) $E\ll m_{\rho}$, GFs can be computed as a
perturbative series in the (small) momentum carried by the current and the
quark masses \cite{Weinberg},\footnote{Throughout this paper we will assume
the chiral limit of QCD, that is we will set the masses of the light quarks
$u$, $d$ and $s$ to zero.} in the so called Chiral Perturbation Theory
($\chi$PT) \cite{Gasser1,Gasser2}, which constitutes a low--energy effective
field theory (EFT). In this formalism the non--perturbative effects are encoded
in the a priori unspecified low--energy constants (LECs) that grow in number
as we go further in the chiral expansion.

At high energies (short distances) the QCD currents in the GF approach each
other and one can expand their product as a tower of local operators in the
Wilson operator product expansion (OPE) \cite{OPEwilson}. In momentum space
this is tantamount to an expansion in inverse powers of momenta, only being
valid in the deep Euclidean region $-\,p^{2}\gg m_{\rho}^{2}$. The vacuum
expectation values of these local operators are known as vacuum condensates
and encode the non--perturbative effects \cite{OPE-ITEP,SVZ2}, which start to
become relevant at intermediate scales. Then we have an explicit separation of
the short distance effects in the Wilson coefficients and the long distance
effects in the condensates. The arbitrariness of this splitting is reflected
in the dependence of both the Wilson coefficients and the condensates on the
renormalisation scale $\mu$ and the employed renormalisation scheme.

At intermediate energies the most reliable tool is an expansion of QCD in
powers of $1/N_{C}$ \cite{largeN1,largeN2,largeN3}, $N_{C}$ being the number
of colours of the QCD gauge group SU$(N_{C})$. This region is populated by
hadronic resonances driving the strong dynamics. At leading order in $1/N_{C}$
there must be an infinite number of such states for any given set of quantum
numbers. GFs of a finite subset of these resonances can be obtained as
tree--level diagrams of an effective Lagrangian with resonances as explicit
degrees of freedom, known as R$\chi$T \cite{RChT1,RChT2}. Nevertheless, we
lack a method to handle an infinite tower of such states, and then the so
called minimal hadronic ansatz (MHA) is usually used, meaning that we only
consider the lowest lying resonances. One is then able to perform a matching
of the three regimes and thus to estimate the values of LECs present in the
$\chi$PT Lagrangian \cite{Moussallam:1997xx,Peris:1998nj}.\footnote{An
approach in the framework of Pad\'e approximants to approximate GFs in the
large-$N_C$ limit which encompasses saturation with a finite number of
resonances, can be found in Ref.~\cite{MP07}.}

The power of EFTs is enhanced by matching onto the more fundamental theory
in a region where both descriptions are sensible, and running down with the
renormalisation group equation (RGE) to a scale where the EFT is useful. This
results in a resummation of large logarithms that would otherwise spoil the
perturbative expansion. As hadronic theories, the matching between $\chi$PT
and R$\chi$T is straightforward since up to a certain mass scale
$\Lambda_{\chi}$ the spectra of both EFTs coincide, and the matching can be
accomplished be formally integrating out the heavy degrees of freedom.
However, the situation changes drastically when matching R$\chi$T onto QCD,
because the spectrum no longer has any particle in common. The reason for it
is that we are going through the chiral phase transition. Then one is forced
to compare GFs computed in the different EFTs in momentum regions where their
domains of validity overlap.

The matching relation is better understood for GFs which are order parameters
of chiral symmetry breaking (order parameters for short). These GFs have a
vanishing Wilson coefficient for the identity operator (that is, the purely
perturbative contribution) to all orders in $\alpha_{s}$ in the chiral limit,
and hence its leading contribution stems from the $\langle\bar{q}q\rangle$
condensate. Since this operator is also responsible for the chiral symmetry
breaking those GFs encode essential information on its mechanism.

This method has been used for two--point GFs including several orders of
$\alpha_{s}$ corrections \cite{mondejar}, but only at leading order for
three--point GFs
\cite{Knecht:2001xc,RuizFemenia:2003hm,Cirigliano:2004ue,SPP,Cirigliano:2006hb}
(see also \cite{Mateu-Portoles} for a matching with two multiplets of
vector--meson resonances). It is the purpose of this paper to push the matching
up to $\mathcal{O}(\alpha_{s})$. Unfortunately, the $\mathcal{O}(\alpha_{s})$
corrections to the $\langle\bar{q}q\rangle$ Wilson coefficient
$C_{\langle\bar{q}q\rangle}$ for three--point GFs are not known, and we will
concentrate on their computation, relegating the details of the matching to a
forthcoming paper. Finally, for the case of the $\Pi_{VT}$ GF, we will show
that we can perform the matching to the one--loop OPE result with a single
multiplet of vector--meson resonances.

The paper is organised as follows: In Section~\ref{sec:mu-dep} we discuss the
renormalisation scale dependence of the GFs and how this affects the matching;
in Section~\ref{sec:2-P} we compute the $\mathcal{O}(\alpha_{s})$ corrections
to the OPE two--point GFs and match the $\langle VT\rangle$ result onto R$\chi$T
with only one multiplet of vector--meson resonances; in Section~\ref{sec:3-P}
we outline the calculation of the $\mathcal{O}(\alpha_{s})$ corrections to the
OPE three--point GFs and argue what the difficulties are when matching onto
the hadronic representation; in Section~\ref{sec:Conclusions} we present our
conclusions.

\section{Scale dependent Green functions\label{sec:mu-dep}}

A strong motivation for computing the $\mathcal{O}(\alpha_{s})$ corrections
to $C_{\langle\bar{q}q\rangle}$ is to become sensitive to the renormalisation
dependence of the quark condensate in full QCD. As is well known, since a QCD
current involves two quark fields in the same space--time point, it is not
enough to renormalise the quark fields to get a finite result. The current
itself must be renormalised in addition:
\begin{equation}
J_\Gamma^B \,\equiv\, Z_\Gamma\,J_\Gamma \,,
\end{equation}
which defines the renormalisation constant $Z_\Gamma$. The superscript $B$
denotes $\mu$--independent bare currents while currents without superscripts
correspond to renormalised (hence generally $\mu$--dependent) ones, where $\mu$
signifies the arbitrary renormalisation scale. It is customary to express the
$\mu$--dependence through the anomalous dimension $\gamma_{\Gamma}$, which is
defined as
\begin{equation}
\label{eq:anom-def-curr}
\gamma_{\Gamma} \,\equiv\, \frac{\mu}{Z_{\Gamma}}\,
\frac{\mathrm{d}Z_{\Gamma}}{\mathrm{d}\mu} \,, \qquad
-\,\mu\,\frac{\mathrm{d}J_{\Gamma}}{\mathrm{d}\mu} \,=\,
\gamma_{\Gamma}\, J_{\Gamma} \,.
\end{equation}
The anomalous dimension depends on the coupling $a_s\equiv\alpha_s/\pi$, and
in perturbation theory has an expansion
\begin{equation}
\gamma_{\Gamma}(a_s) \,=\, \gamma_{\Gamma}^{(1)}\,a_s +
\gamma_{\Gamma}^{(2)}\,a_s^2 + \gamma_{\Gamma}^{(3)}\,a_s^3 + \ldots \,.
\end{equation}

At leading order in $\alpha_{s}$ one can easily calculate the anomalous
dimension of a current with the following master formula (of course, anomalous
dimensions are gauge invariant):
\begin{equation}
\label{eq:master-anomalous}
\gamma_{\Gamma}^{(1)}\,\Gamma \,=\, \frac{C_{F}}{2}\,\Gamma-\frac{C_{F}}{8}\,
\gamma^{\mu}\,\gamma^{\nu}\,\Gamma\,\gamma_{\nu}\,\gamma_{\mu} \,,
\end{equation}
where the algebra should be performed in \emph{four} space--time dimensions.
Plugging in the structures for the different $\Gamma$'s one finds that vector
$V_{\mu}^{a}$ and axial--vector $A_{\mu}^{a}$ currents have zero anomalous
dimension. This is a general result that stems from the fact that in the chiral
limit both currents are conserved. In the $\overline{\mathrm{MS}}$ scheme which
we follow in this paper the renormalisation procedure is mass independent and
so the result for the anomalous dimensions holds for finite quark masses as
well. Similarly, one can show that the quantities $m_{q}\!\cdot\! S^{a}$ and
$m_{q}\!\cdot\! P^{a}$ are $\mu$ independent, if the currents are
normal--ordered. Whence it follows that
\begin{equation}
\label{eq:S-P-m}
\gamma_{S} \,=\, \gamma_{P} \,=\, \frac{\mu}{m_{q}}\,
\frac{\mathrm{d}m_{q}}{\mathrm{d}\mu} \,\equiv\, -\,\gamma_{m} \,,
\end{equation}
to all orders in $\alpha_{s}$. The $\mu$ dependence of the renormalised current
is reflected in the GFs themselves. We can write the dependence through a
RGE~\footnote{This argument is spoiled if the GF needs subtractions, that is
it does not correspond to a physical quantity. Since we are dealing with order
parameters, our GFs do not need any subtraction.}
\begin{equation}
\label{eq:RGE-GF}
\biggl[\, \mu\,\dfrac{\partial}{\partial\mu} + \sum_{i=1}^{n}\gamma_{i}\,
\biggr] \Pi_{1\cdots n} \,=\, 0 \,.
\end{equation}
Here, $\gamma_{i}$ are the anomalous dimensions corresponding to the currents
appearing in $\Pi_{1\cdots n}$. QCD vacuum condensates in general also have a
non--vanishing anomalous dimension, and in particular for the quark condensate
$\langle\bar{q}q\rangle$, as should be clear from the above,
$\gamma_{\langle\bar{q}q\rangle}$ coincides with $\gamma_{S}$ of
Eq.~(\ref{eq:S-P-m}).\footnote{There exists a sophistication in that for a
proper separation of long and short distances in the OPE, non--normal--ordered
condensates should be employed, but then $m\langle\bar qq\rangle$ is not
RG--invariant \cite{Nonqq1,Nonqq2,Nonqq3}. However, here we are working in the
chiral limit, and for vanishing quark masses this problem is absent.}

Let us now consider the case of the 3--point Green function $\Pi_{SSS}$ with
three scalar currents. In the OPE it gets its first contribution from
$\langle\bar{q}q\rangle$, and at leading order its Wilson coefficient is
$\mu$--independent. On the R$\chi$T side the situation is a bit different. It
is well known that each scalar and pseudoscalar current insertion in the chiral
theory is accompanied by a $\langle\bar{q}q\rangle$ factor, so \cite{SPP}
\begin{equation}
\label{eq:OPE-RchiT}
\Pi_{SSS}^{\mathrm{OPE}} \,=\, C_{\langle\bar{q}q\rangle}^{SSS}(p,q)\,
\langle\bar{q}q\rangle(\mu)\,,\qquad\Pi_{SSS}^{\mathrm{R}\chi\mathrm{T}} \,=\,
\widetilde{\Pi}(p,q)\,[\langle\bar{q}q\rangle(\mu)]^3\,,
\end{equation}
and on first sight, we cannot match one onto the other because the $\mu$
dependences appear different. What is happening is that R$\chi$T includes
all $\alpha_{s}$ orders in a non--perturbative fashion, while in QCD the
coefficient function $C_{\langle\bar{q}q\rangle}^{SSS}$ must exhibit some $\mu$
dependence to account for the missing scale factors. This situation easily
generalises to other Green functions, and we can write a RGE for the
coefficient function $C_{\langle\bar{q}q\rangle}(\mu)$, such that
Eq.~(\ref{eq:RGE-GF}) is fulfilled:
\begin{equation}
\label{eq:Wilson-run}
\biggl[\, \mu\,\dfrac{\partial}{\partial\mu} + \gamma_{m} +
\sum_{i=1}^{n}\gamma_{i} \,\biggr] C_{\langle\bar{q}q\rangle}(\mu) \,=\, 0 \,.
\end{equation}
Now the scale dependence of the Green functions in both theories of
Eq.~(\ref{eq:OPE-RchiT}) agrees and in principle the matching could be
performed.

\section{Two--point GFs and matching\label{sec:2-P}}

To prepare the discussion of the 3--point functions below, we begin with
reviewing the calculation of the $\alpha_{s}$ corrections for
$C_{\langle\bar{q}q\rangle}$ in the case of two--point order parameter GFs.
In this simpler scenario, we will discuss the appearance of infrared (IR)
divergences at intermediate stages of the calculation and the renormalisation
of $\langle\bar{q}q\rangle$. For 2--point functions in which the quark
condensate is the leading contribution in the chiral limit, there are only
two GFs to be considered: $\Pi_{AP}^{\mu}$ and $\Pi_{VT}^{\mu,\nu\alpha}$.

The fact that vector and axial--vector currents are conserved in the chiral
limit has implications on the GFs, known as Ward identities. For the case of
the vector--tensor GF they imply that $p_{\mu}\,\Pi_{VT}^{\mu,\nu\alpha}(p)=0$,
and so taking into account the antisymmetry of the tensor current, together
with parity, we can parametrise it as 
\begin{equation}
\left(\Pi_{VT}\right)_{\mu,\nu\rho}^{ab}(p) \,=\, i\,\delta^{ab}\,
( g_{\mu\nu}\,p_{\rho} - g_{\mu\rho}\,p_{\nu} )\,\Pi_{VT}(p^{2}) \,.
\end{equation}
For the $\Pi_{AP}^{\mu}$ GF, Ward identities have somewhat deeper consequences
and in fact completely fix it in the chiral limit:
\begin{equation}
\label{eq:pion-satur}
\left(\Pi_{AP}\right)_{\mu}^{ab}(p) \,=\, 2\, i\,\delta^{ab}\,
\dfrac{p_{\mu}}{p^{2}}\, \left\langle \bar{q}q\right\rangle \,,
\end{equation}
where a detailed derivation can be found in Appendix~A. Again, we see that the
scale dependencies on both sides of Eq.~(\ref{eq:pion-satur}) are identical
since the anomalous dimensions of the scalar current $\gamma_S$ and of the
quark condensate agree. Furthermore, one finds that in the chiral limit
$\Pi_{AP}$ is saturated by one pion exchange.

\FIGURE[thb]{\includegraphics[angle=0, width=13cm]{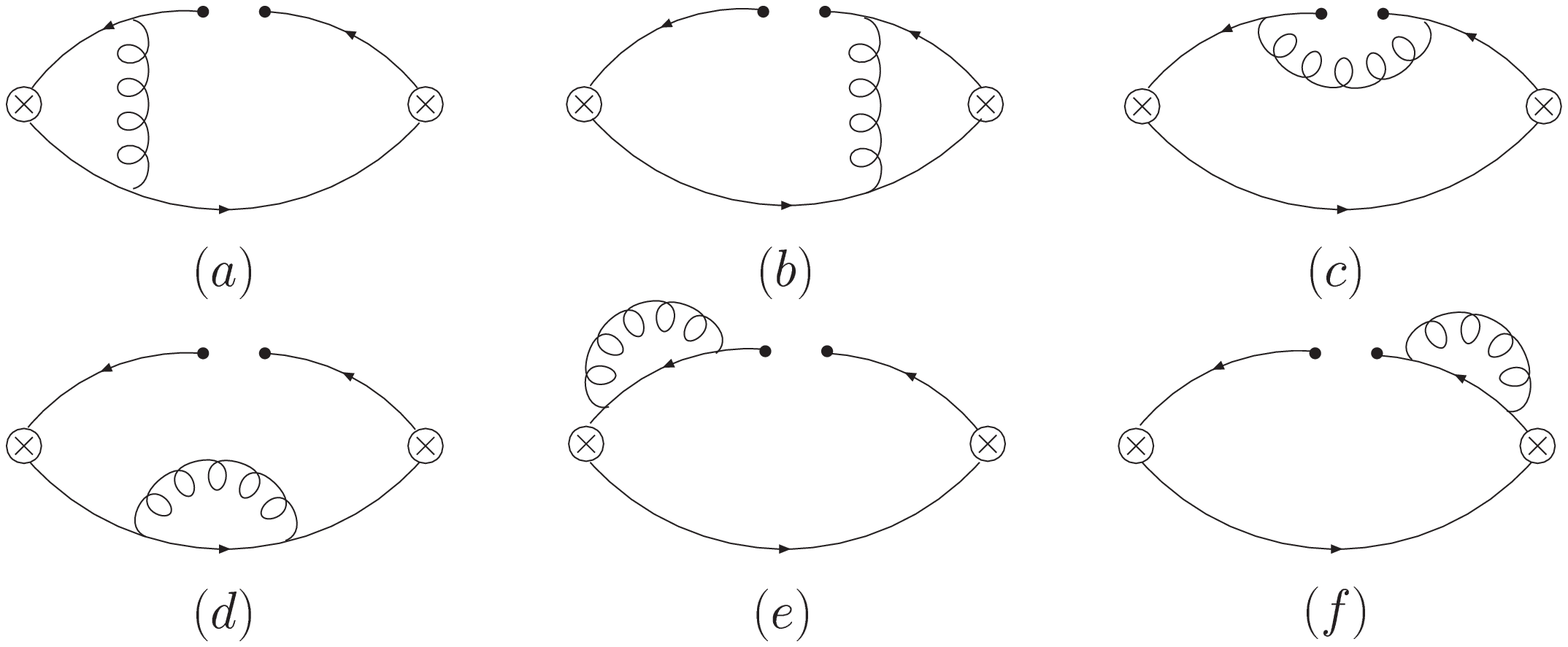}
\caption{Gluonic corrections to the $C_{\langle\bar{q}q\rangle}$ for two--point
GFs. Diagrams (c), (e) and (f) are infrared divergent.\label{fig:2-P_GF}}}

Let us now concentrate on the $\Pi_{VT}^{\mu,\nu\alpha}$ GF. The relevant
diagrams are shown in Fig.~\ref{fig:2-P_GF}. Diagrams (c), (e) and (f) are
infrared divergent since they involve gluons attached to quark lines with
zero momentum. The same diagrams contribute to $\Pi_{AP}$ and
Eq.~(\ref{eq:pion-satur}) shows that it is free from IR divergences. So they
cancel when adding the six diagrams, and the same occurs for the $\Pi_{VT}$ GF.
Since these divergences cancel at the end we might choose any method to
regularise them. In what follows, we will adopt dimensional regularisation,
which simplifies computations notoriously. In this scheme diagrams (e) and (f)
are zero: they are scaleless and convert the IR divergence of diagram (c)
into an ultraviolet (UV) one. In particular, for this diagram we obtain
\begin{equation}
\Pi_{VT}^{(c)} \,=\, a_s\,\frac{C_{F}}{4}\,\frac{\langle\bar{q}q\rangle}{p^{2}}
\biggl\{\, -\,(3+a)\biggl[\frac{1}{\hat{\epsilon}}-\log\left(-\,\frac{p^{2}}
{\mu^{2}}\right)\biggr] - 1 - a \,\biggr\} \,,
\end{equation}
where $a$ denotes the gauge parameter.\footnote{All calculations in this paper
have been performed in an arbitrary covariant gauge. The dependence on the
gauge parameter $a$ cancels in all  our final, physical results, constituting
a good check for them.} As we see, by itself this diagram is still gauge
dependent, which means that the other diagrams are required in order to obtain
a gauge--invariant result.

Summing up all diagrams, we find a remaining divergence of the form
$-\,C_F\, a_s/\hat{\epsilon}$. Part of this divergence (namely
$-\,3\,C_F\, a_s/4\,\hat{\epsilon}$) is absorbed in the renormalisation of the
quark condensate, that is $\langle\bar qq\rangle^B=Z_{\langle\bar qq\rangle}
\langle\bar qq\rangle(\mu)$, while the remaining divergence is removed with
the renormalisation of the tensor current, $J_T^B=Z_T J_T(\mu)$, leading to
\begin{equation}
\label{ZT}
Z_T \,=\, 1 - \frac{C_F}{4}\,a_s\,\frac{1}{\hat{\epsilon}} + {\cal O}(a_s^2)\,.
\end{equation}
From the renormalisation constant $Z_T$, we can also derive the first
coefficient of the tensor--current anomalous dimension, $\gamma_T^{(1)}=C_F/2$,
in agreement with the simple formula of Eq~(\ref{eq:master-anomalous}). For
the full vector--tensor Green function at the next--to--leading order, we then
find
\begin{equation}
\label{eq:VT-OPE}
\Pi_{VT}^{\mathrm{OPE}}(p^{2},\mu) \,=\,
\frac{\left\langle \bar{q}q\right\rangle (\mu)}{p^{2}}\left\{ 1 +
a_s\, C_{F}\left[\log\left(-\,\frac{p^{2}}{\mu^{2}}\right) - 1 \right]\right\}
+ \mathcal{O}(a_s^2,p^{-4}) \,,
\end{equation}
which is of course gauge invariant. For $N_{C}$ flavours
$C_{F}=\frac{N_{C}^{2}-1}{{2N}_{C}}\approx\frac{N_{C}}{2}$ where the
approximation corresponds to the large--$N_{C}$ limit. With our result of
Eq.~(\ref{eq:VT-OPE}), we can check that the RGE for the coefficient function
is trivially satisfied:
\begin{equation}
\label{eq:T-RGE}
\left[\, \mu\,\dfrac{\partial}{\partial\mu} + \gamma_{m} +\gamma_{T} \,\right]
C_{\left\langle \bar{q}q\right\rangle }^{VT} \,=\, \left[\, \mu\,
\dfrac{\partial}{\partial\mu} + 2\,a_s C_{F} \,\right]
C_{\left\langle \bar{q}q\right\rangle }^{VT} \,=\, 0 \,.
\end{equation}

Let us now match this result onto the large--$N_{C}$ prediction of R$\chi$T.
Since the $\langle VT\rangle$ GF in the chiral limit does not need to be
subtracted, it is completely determined by its spectral function. Then it can
be regarded as an observable and thus we can directly match its expression in
different approximations, such as OPE and R$\chi$T. However, we still have the
problem that the tensor current requires renormalisation, and thus, the
$\langle VT\rangle$ GF is scale and scheme dependent. This renormalisation
dependence would then be reflected in a scale dependent coupling of the tensor
current to vector mesons $f_V^T(\mu)$ on the hadronic side. Since we prefer
to work with hadronic quantities which are explicitly scale independent,
another possibility is to multiply $f_V^T(\mu)$ by an appropriate scale
factor $R_T(\mu)$, which results in a scale independent tensor decay constant
$\hat f_V^T$. This is analogous to the definition of scale--invariant
$B$--factors, which parametrise hadronic matrix elements of four--quark
operators, in the case of weak hadronic decays \cite{BJW90}. Therefore, we
define
\begin{eqnarray}
\label{FThat}
\hat f_V^T &\,\equiv\,& f_V^T(\mu)\,R_T(\mu) \,\equiv\, f_V^T(\mu)\,
\exp\biggl\{-\!\!\int\limits^{a_s(\mu)}\!\frac{\gamma_T(a_s)}{\beta(a_s)}\,
{\rm d}a_s \biggr\} \nonumber \\
\vbox{\vskip 8mm}
&\,=\,& f_V^T(\mu)\,[a_s(\mu)]^{-\gamma_T^{(1)}/\beta_1} \biggl[\, 1 -
\biggl( \frac{\gamma_T^{(2)}}{\beta_1} - \frac{\beta_2\,\gamma_T^{(1)}}
{\beta_1^2} \biggr) a_s(\mu) + {\cal O}(a_s^2) \,\biggr] \\
\vbox{\vskip 8mm}
&\,\stackrel{N_f=3}{=}\,& f_V^T(\mu)\, [a_s(\mu)]^{-\frac{4}{27}} \biggl[\,
1 - \frac{337}{486}\,a_s(\mu) + {\cal O}(a_s^2) \,\biggl] \,. \nonumber
\end{eqnarray}
The anomalous dimension of the tensor current is known up to order $a_s^3$
\cite{BG95,Gra00}, and thus one could even extend Eq.~(\ref{FThat}). However,
at the order considered here this does not make sense, since we only stay at
the next--to--leading order level. Now multiplying our result (\ref{eq:VT-OPE})
for the Green function with the scale factor $R_T(\mu)$, it is a trivial
exercise to convince oneself that
$R_T(\mu)\,\Pi_{VT}^{\mathrm{OPE}}(p^{2},\mu)$ is scale independent at the
considered order. Nevertheless, it should be kept in mind that it still
depends on the renormalisation scheme, for example on the scheme in which
$\langle\bar qq\rangle$ is renormalised.

In principle, since the next--to--leading order result for the
$\langle VT\rangle$ GF contains a logarithm in the dynamical variable $p^2$,
let us strongly emphasise that an infinite tower of resonances would be
required for a sound matching to the R$\chi$T. Still, as a simple--minded
approach, we will next consider the aforementioned minimal hadronic ansatz
(MHA). This amounts to the assumption that a single resonance is enough to
correctly describe the physics in a certain energy regime.\footnote{For a
discussion of the subtleties arising when matching the full tower of
vector--meson resonances to the OPE for the $\langle VT\rangle$ GF the reader
is referred to Ref.~\cite{Cata:2008zc}.} Tensor sources in chiral Lagrangians
were first introduced in Refs.~\cite{Cata-Mateu-1}. In
Ref.~\cite{Mateu-Portoles} the matching for the $\langle VT\rangle$ GF was
performed at $\mathcal{O}(\alpha_s^0)$, and here we will include the
$\mathcal{O}(\alpha_s)$ corrections. Again multiplying the GF with the scale
factor in order to obtain a scale--invariant quantity, the hadronic ansatz
reads:
\begin{equation}
\label{eq:MHA-VT}
R_T(\mu)\,\Pi_{VT}^{\mathrm{R}\chi\mathrm{T}}(p^{2}) \,=\,
2\,\frac{ f_V\, \hat f_V^T\, m_{V}}{m_{V}^{2}-p^{2}} \,.
\end{equation}
The precise definitions for the decay constants $f_V$ and $f_V^T$ can be found
in Eq.~(53) of Ref.~\cite{Mateu-Portoles}. Eq.~(\ref{eq:MHA-VT}) is in
principle assumed to be valid at all energies at leading order in $1/N_C$,
since it incorporates chiral symmetry and the correct high--energy behaviour.
It can be expanded in inverse powers of $p^2$, permitting a direct comparison
with the OPE in Eq.~(\ref{eq:VT-OPE}).  However while Eq.~(\ref{eq:MHA-VT}) is
explicitly scale independent, (\ref{eq:VT-OPE}) contains a logarithm which
compensate the running of the tensor source and the quark condensate.

To perform the matching in practice, we choose a particular matching point
and scale. First of all, to sum up the logarithm, we will employ the scale
$\mu^2=-\,p^2\equiv M^2$. Then, $M^2$ should be large enough so that only
keeping the first term in the OPE is a good approximation, while it should not
be too large so that only putting one resonance on the hadronic side is
reasonable. From these considerations, we would conclude, that $M$ should be
in the range $1$--$2\,$GeV. For the matching relation, we then find
\begin{equation}
\label{eq:matching}
f_V\,\hat f_V^T\, m_{V} \,=\, -\,2\,[a_s(M^2)]^{-\frac{4}{27}
\left\{\frac{3}{22}\right\}}
\langle\bar qq\rangle(M^2)\left[\, 1 - \frac{985}{486}
\left\{\frac{8357}{11616}\,N_C\right\}\,a_s(M^2) \,\right] \,,
\end{equation}
where in the curly brackets, we have also included the numbers corresponding
to the large--$N_C$ limit. Eq.~(\ref{eq:matching}) can be viewed as a refinement
over the analogous estimate of Ref.~\cite{Mateu-Portoles}.

Let us finally come to a numerical analysis of Eq.~(\ref{eq:matching}).
Employing the central values $f_V = 221\,$MeV and $M_V = 775\,$MeV
\cite{Mateu-Portoles}, as well as the value for the quark condensate
$\langle\bar qq\rangle(2\,{\rm GeV})=-\,(267\,{\rm MeV})^3$ \cite{Nonqq3},
and $\alpha_s(M_Z)=0.119$, we obtain
\begin{equation}
\label{eq:FVTnum}
\hat f_V^T \,=\, 195 \pm 55 \;{\rm MeV} \,,
\end{equation}
where the quoted uncertainty dominantly results from a variation of the
matching scale $M$ in the range $1$--$2\,$GeV, and to a lesser extent from
either taking the renormalisation--group coefficients in full QCD, or the
large--$N_C$ limit. The large matching--scale dependence of our result reflects
the imperfection of the matching. At a scale of $1\,$GeV, the scale dependent
vector--meson tensor coupling reads: $f_V^T(1\,{\rm GeV}) = 167\pm 47\,$MeV.
Given the large uncertainties from the matching scale, this finding is in
surprising agreement to the leading order result
$f_V^T(1\,{\rm GeV}) = 165\,$MeV \cite{Mateu-Portoles} and to determinations
of the tensor coupling $f_V^T$ from QCD sum rules and lattice QCD of
Refs.~\cite{BB96,Braunetal03}.

\section{Three--point GFs\label{sec:3-P}}

Let us now concentrate on the main topic of this work: the $\alpha_{s}$
corrections to $C_{\langle\bar{q}q\rangle}$ for the three--point
order--parameter GFs.  Three--point GFs are of great interest because of
several reasons. First, unlike the two--point case, there is a quite large
amount of them that are order parameters of the chiral symmetry breaking.
Second, in the framework of R$\chi$T they involve vertices between resonances
and so they are useful for studying how the resonances interact. Third, there
are many $\mathcal{O}(p^{6})$ $\chi$PT LECs that can be determined with these
GFs \cite{Cirigliano:2006hb}. And fourth, by means of the LSZ reduction formula
we can relate the GFs with form factors entering the calculation of many
interesting hadronic observables.  Some phenomenological applications of these
Green functions can be found in Ref.~\cite{Narison1,Narison2} and references
therein. 

There are eight such GFs and we will sort them according to their
intrinsic--parity and Lorentz tensor rank. For rank--two GFs we can use the
chiral Ward identities and the transformation properties under parity for
splitting them in terms of only one or two Lorentz tensors (made up of
$g^{\mu\nu}$, $\varepsilon^{\mu\nu\alpha\beta}$~\footnote{In this work we use
the convention $\varepsilon^{0123}=+1$.} and the external momenta). This
amounts to a great simplification on their calculation. When two or more
currents have the same Dirac structure one can also exploit Bose symmetry.
We already saw that the flavour structure of two--point GFs is trivial: it is
proportional to a Kronecker delta (there is no other SU$(3)$ rank--two
tensor). For the present case there are only two rank--three SU$(3)$ tensors,
the totally antisymmetric structure constant $f^{abc}$ and the totally
symmetric $d^{abc}$ tensor. Either one or the other will appear as a global
factor depending on the transformation properties of the GF under time
reversal.

\FIGURE[htb]{\includegraphics[angle=0, width=13cm]{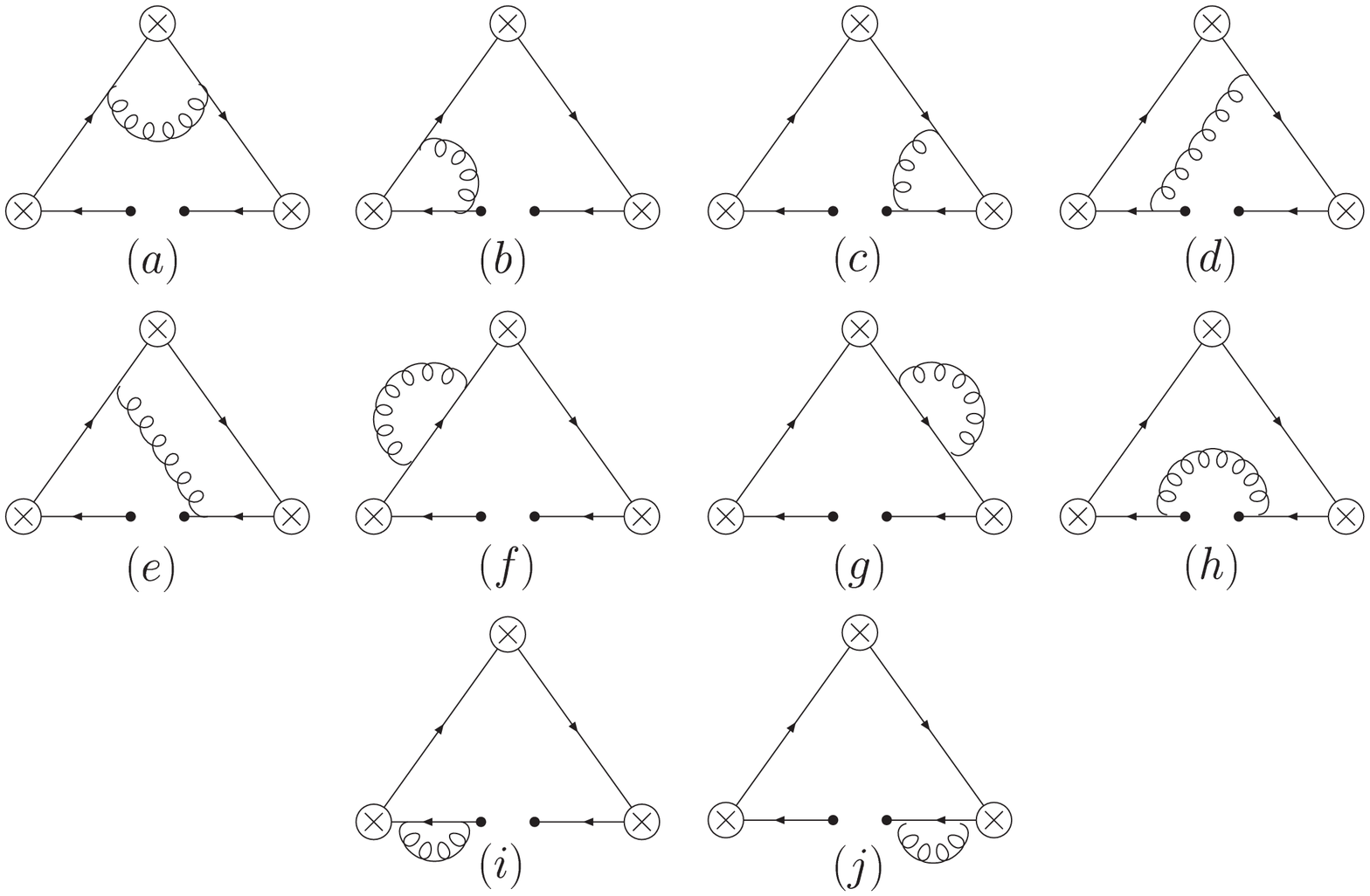}
\caption{Gluonic corrections to $C_{\langle\bar{q}q\rangle}$ for three point
GFs.\label{fig:3-P}}}

The diagrams giving rise to the gluonic corrections are shown in
Fig.~\ref{fig:3-P}. Diagrams (h), (i) and (j) are analogous to those in
Fig.~\ref{fig:2-P_GF} (c), (e) and (f), respectively. Diagrams (d) and (e) are
IR and UV safe, even though they have gluons attached to zero momentum quark
lines. Much as we did in Section~\ref{sec:2-P}, we regularise those divergences
in dimensional regularisation, renormalise the quark condensate as in the last
section and after summing up all diagrams IR divergences will cancel out. The
rest of the (UV) divergences are absorbed in counterterms as usual, resulting
in a finite (but scale dependent) result.

Loop corrections manifest themselves as logarithms, dilogarithms and constant
pieces. In general we will have the following decomposition:
\begin{equation}
\label{eq:3-P-C-gen}
C_{\langle\bar{q}q\rangle}^{\alpha_{s}} \,=\, a_s\,\frac{C_{F}}{8}
\left[ L_{p}\log\left(-\dfrac{p^{2}}{\mu^{2}}\right) +
L_{q}\log\left(-\dfrac{q^{2}}{\mu^{2}}\right) +
L_{r}\log\left(-\dfrac{r^{2}}{\mu^{2}}\right) + L_{d}\, C_{0} + L_{c}\right] ,
\end{equation}
where $r\equiv -\,(p+q)$ and the $L_{i}$ are $\mu$--independent meromorphic
functions of the squared external momenta. This simple structure arises because
in the chiral limit all internal lines, either quark or gluon, are massless.
The massless, scalar three--point integral $C_{0}(p^{2},q^{2},r^{2})$ collects
all dilogarithms and its explicit expression reads \cite{tHV79}:
\begin{eqnarray}
C_{o}\left(p^{2},q^{2},r^{2}\right) &\,\equiv\,& -\,i\,(4\pi)^2\!
\int\!\frac{\mathrm{d}^4\!k}{(2\pi)^4}\,\frac{1}{k^2(p+k)^2(q-k)^2} \nonumber \\
\vbox{\vskip 8mm}
&\,=\,& \frac{1}{\sqrt{\lambda}}\,\left\{ \mathrm{Li}_{2}\left(-\frac{\lambda+q^{2}+p^{2}-r^{2}}{\lambda-q^{2}-p^{2}+r^{2}}\right)-\,\mathrm{Li}_{2}\left(-\frac{\lambda-q^{2}-p^{2}+r^{2}}{\lambda+q^{2}+p^{2}-r^{2}}\right)\right.\nonumber \\
\vbox{\vskip 6mm}
&& \hspace{6.8mm}+\,\mathrm{Li}_{2}\left(-\frac{\lambda+q^{2}+r^{2}-p^{2}}{\lambda-q^{2}-r^{2}+p^{2}}\right)-\,\mathrm{Li}_{2}\left(-\frac{\lambda-q^{2}-r^{2}+p^{2}}{\lambda+q^{2}+r^{2}-p^{2}}\right)\nonumber \\
\vbox{\vskip 6mm}
&& \hspace{6.4mm}+\left.\mathrm{Li}_{2}\left(-\frac{\lambda+r^{2}+p^{2}-q^{2}}{\lambda-r^{2}-p^{2}+q^{2}}\right)-\,\mathrm{Li}_{2}\left(-\frac{\lambda-r^{2}-p^{2}+q^{2}}{\lambda+r^{2}+p^{2}-q^{2}}\right)\right\} ,
\end{eqnarray}
where $\lambda$ is the well--known K\"allen function $\lambda\left(p^{2},q^{2},r^{2}\right)=\left(p^{2}+q^{2}-r^{2}\right)^{2}-4\, p^{2}q^{2}$.
Up to order $\alpha_s$, the $\mu$ dependence of $C_{\langle\bar{q}q\rangle}$
then corresponds to
\begin{equation}
-\,\mu\,\dfrac{\mathrm{d}}{\mathrm{d}\mu}\,C_{\langle\bar{q}q\rangle}(\mu)
\,=\, a_s \frac{C_{F}}{4}\,(L_{p}+L_{q}+L_{r}) \,.
\end{equation}
From the general expression Eq.~(\ref{eq:3-P-C-gen}) we see that there is no
possible choice of $\mu$ that makes all logarithms cancel simultaneously. Then
we are forced to include an infinite number of resonances in order to match
onto R$\chi$T. Moreover, besides logarithms we also have the
($\mu$--independent) dilogarithms, and it is unclear how to match those to the
sum of resonance contributions. Let us next explore the different sectors.

\subsection{Zero--rank GFs: $\langle$SSS$\rangle$ and
$\langle$SPP$\rangle$\label{sec:SSS} }

Before starting the discussion of this sector we remind the reader that GFs of
the type $\langle SSP\rangle$ or $\langle PPP\rangle$ are forbidden by parity
invariance of the strong interactions. The Lorentz structure of this sector is
trivial, as it has no Lorentz index; time reversal symmetry fully determines
the flavour structure:
\begin{equation}
\Pi_{SSS(SPP)}^{abc}(p^{2},q^{2},r^{2})\,=\, d^{abc}\,\Pi_{SSS(SPP)}(p^{2},q^{2},r^{2})\,,
\end{equation}
Bose symmetry requires $\Pi_{SSS}(p^{2},q^{2},r^{2})$ to be totally symmetric
in its three arguments and also
$\Pi_{SPP}(p^{2},q^{2},r^{2})=\Pi_{SPP}(p^{2},r^{2},q^{2})$.

The total anomalous dimension of these GFs is
$\gamma=3\,\gamma_{S}=-\,3\,\gamma_{m}$, as discussed in
Section~\ref{sec:mu-dep}. Thus when calculating them in R$\chi$T (or $\chi$PT)
one finds that they are proportional to
$\langle\bar{q}q\rangle^{3}$.\footnote{In $\chi$PT with its hadronic degrees
of freedom, there is no dynamical generation of the QCD renormalisation scale
$\mu$ (\emph{i.e.} it does not come from the UV divergences of Feynman diagrams)
and thus it is fully contained in $B_{0}=-\,\langle\bar{q}q\rangle/F^{2}$, the
quark masses and unphysical couplings like $H_1^r$ and $H_2^r$. Furthermore,
all LEC of the additional operators in the chiral Lagrangian which arise in
the presence of tensor sources \cite{Cata-Mateu-1} also depend on $\mu$. 
The QCD renormalisation scale $\mu$ is not to be confused with the chiral
renormalisation scale $\mu_{\chi}$, showing up when computing chiral
logarithms.} On the other hand in the OPE side the first non--vanishing
contribution comes from the (single) quark condensate. The RGE for its Wilson
coefficient, Eq.~(\ref{eq:Wilson-run}) is then 
\begin{equation}
\left( \mu\,\dfrac{\partial}{\partial\mu}\,-2\,\gamma_{m} \right)
C_{\langle\bar{q}q\rangle}^{SSS(SPP)}(\mu) \,=\, 0 \,,
\end{equation}
where the Wilson coefficients $C_{\langle\bar{q}q\rangle}^{SSS(SPP)}(\mu)$
have been defined as
\begin{equation}
\Pi_{SSS(SPP)}(p^{2},q^{2},r^{2},\mu)\,=\, C_{\langle\bar{q}q\rangle}^{SSS(SPP)}(p^{2},q^{2},r^{2},\mu)\langle\bar{q}q\rangle(\mu)\,.
\end{equation}
At $\mathcal{O}(\alpha_{s}^{0})$ their expressions read:
\begin{equation}
\label{eq:CSSSLO}
C_{\langle\bar{q}q\rangle}^{SSS}\,=\,-\,\frac{2}{p^{2}\,q^{2}\,r^{2}}\,\lambda\left(p^{2},q^{2},r^{2}\right)\,,\qquad C_{\langle\bar{q}q\rangle}^{SPP}\,=\,\frac{2}{p^{2}\,q^{2}\,r^{2}}\left[p^{4}-\left(q^{2}-r^{2}\right)^{2}\right]\,.
\end{equation}
Our explicit results for the $\mathcal{O}(\alpha_{s})$ corrections to these
expressions (and for the remaining Wilson coefficients) are rather lengthy and
thus have been relegated to Appendix~B.

In Ref.~\cite{SPP}, the leading order expressions (\ref{eq:CSSSLO}) have been
matched to hadronic representations which took into account constraints
arising from $\chi$PT and R$\chi$T, thereby allowing to estimate some of the
LECs in these frameworks. Going to the next--to--leading order, now we have to
face two complications. First of all, we have the question of scale dependence
of the GFs. This problem could be treated like the $\langle VT\rangle$ GF of
the last section: by defining quantities which are explicitly scale independent
through multiplication with appropriate scale factors. (This problem cannot
even be addressed at the leading order.) Secondly, due to the presence of
logarithms (and dilogarithms) the matching definitely requires an infinite
tower of resonances, making it much more involved. For this reason, we postpone
a detailed discussion of the matching procedure for three--point
order--parameter GFs to a forthcoming publication.\footnote{An additional
complication in satisfying quark--counting rules that requires the matching
of higher--point functions with an infinite set of resonances is discussed in
Ref.~\cite{BGLP03}.}

\subsection{Odd--intrinsic--parity sector: $\langle$VVP$\rangle$,
$\langle$AAP$\rangle$ and $\langle$VAS$\rangle$ GFs
\label{sec:VVP}}

These GFs are rank--two Lorentz tensors and so in principle they can be
written in terms of many independent Lorentz tensors built from external
momenta, the metric tensor $g^{\mu\nu}$ and the Levi--Civita symbol
$\varepsilon^{\mu\nu\alpha\beta}$. However chiral Ward identities imply
$p_{\mu}\,\Pi_{123}^{\mu\nu}=q_{\nu}\,\Pi_{123}^{\mu\nu}=0$. Besides, the
transformation properties of the odd--intrinsic--parity sector can only
accommodate a Levi--Civita symbol contracted with the two external momenta.
Together with time reversal invariance we are left with the following structure:
\begin{equation}
\left(\Pi_{VV\!P(AAP)[V\!AS]}\right)_{\mu\nu}^{abc}(p,q)\,=\,\Pi_{VV\!P(AAP)[V\!AS]}(p^{2},q^{2},r^{2})\,\varepsilon_{\mu\nu\alpha\beta}\, p^{\alpha}q^{\beta}d^{abc}(d^{abc})[f^{abc}]\,.
\end{equation}
Bose symmetry implies
$\Pi_{VV\!P(AAP)}(p^{2},q^{2},r^{2})=\Pi_{VV\!P(AAP)}(q^{2},p^{2},r^{2})$.
The total anomalous dimension of these currents is
$\gamma=\gamma_{S}=-\,\gamma_{m}$, and this means that both in R$\chi$P
($\chi$PT) and leading order OPE the result is proportional to a single quark
condensate. Moreover, the RGE for the Wilson coefficient
Eq.~(\ref{eq:Wilson-run}) reduces to the fact that it is $\mu$--independent:
\begin{equation}
\dfrac{\mathrm{d}}{\mathrm{d}\mu}\, C_{\langle\bar{q}q\rangle}^{VV\!P(AAP)[V\!AS]}\,=\,0\,.
\end{equation}
Analogously to the $\langle SSS\rangle$ and $\langle SPP\rangle$ GFs, we
then define the Wilson coefficients as 
\begin{equation}
\Pi_{VV\!P(AAP)[V\!AS]}(p^{2},q^{2},r^{2},\mu)\,=\, C_{\langle\bar{q}q\rangle}^{VV\!P(AAP)[V\!AS]}(p^{2},q^{2},r^{2})\,\langle\bar{q}q\rangle(\mu)\,.
\end{equation}

At lowest order they are given by
\begin{equation}
C_{\langle\bar{q}q\rangle}^{VV\!P} \,=\,
\frac{p^{2}+q^{2}+r^{2}}{q^{2}\,p^{2}\,r^{2}} \,, \quad
C_{\langle\bar{q}q\rangle}^{AAP} \,=\,
\frac{p^{2}+q^{2}-r^{2}}{q^{2}\,p^{2}\,r^{2}} \,, \quad
C_{\langle\bar{q}q\rangle}^{V\!AS} \,=\,
\frac{p^{2}-q^{2}-r^{2}}{q^{2}\,p^{2}\,r^{2}} \,.
\end{equation}
The next--to--leading order $\alpha_s$ corrections are again presented in
Appendix~B. Here a technical comment is in order. Since we work in dimensional
regularisation, there is a question how to treat $\gamma_5$. In all our
computations we have employed a fully anticommuting $\gamma_5$. Either two
$\gamma_5$'s appear in a trace and can be cancelled before taking the trace,
or, in the odd--parity sector, we can first perform the $\gamma$ contractions
before taking the trace, and are then left with traces of only four
$\gamma$--matrices and a $\gamma_5$ which are unambiguous. As expected, in the
odd--parity sector our Wilson coefficients turn out to be scale independent.

\subsection{Even--intrinsic--parity sector: $\langle$VVS$\rangle$,
$\langle$AAS$\rangle$ and $\langle$VAP$\rangle$ GFs
\label{sec:VAP}}

These last GFs are also rank--two tensors and so they can be built out of many
Lorentz structures as well. The chiral Ward identities are not so simple as in
the odd--intrinsic--parity sector and read:
\begin{eqnarray}
\label{eq:VAP}
p^{\mu}\,(q^{\nu})\left(\Pi_{VV\!S}\right)_{\mu\nu}^{abc}(p,q) &\,=\,& 0 \,,
\nonumber \\
\vbox{\vskip 6mm}
p^{\mu}\,(q^{\nu})\left(\Pi_{AAS}\right)_{\mu\nu}^{abc}(p,q) &\,=\,&
2\,\langle\bar{q}q\rangle\, d^{abc}\,\frac{q_{\nu}}{q^2}\,\biggl(
\frac{p_{\mu}}{p^2}\biggr)\,, \nonumber \\
\vbox{\vskip 6mm}
p^{\mu}\left(\Pi_{V\!AP}\right)_{\mu\nu}^{abc}(p,q) &\,=\,&
-\,2\,\langle\bar{q}q\rangle\, f^{abc} \biggl( \frac{q_\nu}{q^{2}} +
\frac{r_\nu}{r^2} \biggr) \,, \nonumber \\
\vbox{\vskip 6mm}
q^{\nu}\left(\Pi_{V\!AP}\right)_{\mu\nu}^{abc}(p,q) &\,=\,&
2\,\langle\bar{q}q\rangle\, f^{abc}\,\frac{r_\mu}{r^2} \,,
\end{eqnarray}
where we have also used time reversal invariance. These results rely on the
fact that the $\langle AP\rangle$ GF is fully determined by the chiral Ward
identity Eq.~(\ref{eq:pion-satur}), and they must be satisfied in any sensible
description of the strong interactions. For instance, concentrating on
$\langle AAP\rangle$ and $\langle V\!AS\rangle$, in R$\chi$T this means that
once we contract with one momenta, only the pion pole can survive. In the OPE
they also have deep implications: since there is no other condensate than
$\langle\bar{q}q\rangle$ in Eq.~(\ref{eq:VAP}), it implies that the contribution
of higher dimension condensates vanishes when contracting with one external
momenta, and since there is no $\alpha_{s}$ factor all contributions to the
$\langle\bar{q}q\rangle$ beyond leading order must also vanish when the
contraction is performed. Eq.~(\ref{eq:VAP}) allows us to write the GFs as
\begin{eqnarray}
\left(\Pi_{VVS}^{\mu\nu}\right)^{abc}\left(q^{2},p^{2},r^{2}\right) &\,=\,&
d^{abc}\Big[\, P^{\mu\nu}(p,q)\,\mathcal{F}_{VVS}\!\left(p^{2},q^{2},r^{2}
\right) + Q^{\mu\nu}(p,q)\,\mathcal{G}_{VVS}\left(p^{2},q^{2},r^{2}\right)
\,\Big] , \nonumber \\
\vbox{\vskip 8mm}
\left(\Pi_{AAS}^{\mu\nu}\right)^{abc}\left(q^{2},p^{2},r^{2}\right) &\,=\,&
f^{abc}\biggl[\, P^{\mu\nu}(p,q)\mathcal{\, F}_{AAS}\!\left(p^{2},q^{2},r^{2}
\right) + Q^{\mu\nu}(p,q)\,\mathcal{G}_{AAS}\!\left(p^{2},q^{2},r^{2}\right)
\nonumber \\
\vbox{\vskip 6mm}
&& \hspace{1cm} +\, 2\left\langle \bar{q}q\right\rangle \,
\frac{p^{\mu}\,q^{\nu}}{q^{2}\,p^{2}} \,\biggr] \,, \nonumber \\
\vbox{\vskip 8mm}
\left(\Pi_{V\!AP}^{\mu\nu}\right)^{abc}\left(q^{2},p^{2},r^{2}\right) &\,=\,&
f^{abc}\biggl[\, P^{\mu\nu}(p,q)\,\mathcal{F}_{V\!AP}\!\left(p^{2},q^{2},
r^{2}\right) + Q^{\mu\nu}(p,q)\,\mathcal{G}_{V\!AP}\!\left(p^{2},q^{2},r^{2}
\right) \nonumber \\
\vbox{\vskip 6mm}
&& \hspace{1cm} -\,2\left\langle \bar{q}q\right\rangle \!\biggl(
\frac{\left(p+2\,q\right)^{\mu}q^{\nu}}{q^{2}\,r^{2}}-\frac{g^{\mu\nu}}{r^{2}}
\biggr) \,\biggr] \,, \label{eq:VAP-param}
\end{eqnarray}
where the transverse tensors $P^{\mu\nu}$ and $Q^{\mu\nu}$ are defined as
\begin{eqnarray}
P^{\mu\nu}(p,q) &\,\equiv\,& q^{\mu}\,p^{\nu}-(p\cdot q)\, g^{\mu\nu}\,,
\nonumber \\
\vbox{\vskip 6mm}
Q^{\mu\nu}(p,q) &\,\equiv\,& p^{2}\, q^{\mu}\,q^{\nu}+q^{2}\, p^{\mu}\,p^{\nu}
- (p\cdot q)\, p^{\mu}\,q^{\nu}-p^{2}\,q^{2}\, g^{\mu\nu} \,.
\end{eqnarray}
$\mathcal{F}$ and $\mathcal{G}$ are two scalar functions satisfying Bose
symmetry in the first two arguments for the $\langle VV\!S\rangle$ and
$\langle AAS\rangle$ GFs:
\begin{equation}
{\mathcal{F}\,[\mathcal{G}]}_{\,VV\!S\,(AAS)}\left(p^{2},q^{2},r^{2}\right)\,=\,
{\mathcal{F}\,[\mathcal{G}]}_{\,VV\!S\,(AAS)}\left(q^{2},p^{2},r^{2}\right) \,.
\end{equation}

At lowest order the determination of the scalar functions $\mathcal{F}$ and
$\mathcal{G}$ is straightforward, but once we go beyond this level their direct
computation turns out to be rather complicated. Instead it will be much simpler
to concentrate on the determination of linear combinations of these factors,
obtained by taking appropriate contractions in Eq.~(\ref{eq:VAP-param}) (i.e.
those contractions that do not reduce to a Ward identity of Eq.~(\ref{eq:VAP})).
They take the form
\begin{eqnarray}
g_{\mu\nu}\,\Pi_{VV\!S}^{\mu\nu} &\,=\,&
\frac{3}{2}\left(p^{2}+q^{2}-r^{2}\right)\mathcal{F}_{VV\!S} -
\Big(\frac{\lambda}{4} + 3\,p^{2} q^{2}\Big)\mathcal{G}_{VV\!S} \,,\nonumber \\
\vbox{\vskip 6mm}
q_{\mu}\,p_{\nu}\Pi_{VV\!S}^{\mu\nu} &\,=\,&
-\,\frac{\lambda}{4}\,\mathcal{F}_{VV\!S} + \frac{\lambda}{8}\left(p^{2}+
q^{2}-r^{2}\right)\mathcal{G}_{VV\!S} \,, \nonumber \\
\vbox{\vskip 6mm}
g_{\mu\nu}\,\Pi_{AAS}^{\mu\nu} &\,=\,&
\frac{3}{2}\left(p^{2}+q^{2}-r^{2}\right)\mathcal{F}_{AAS} -
\Big(\frac{\lambda}{4} + 3\,p^{2} q^{2}\Big)\mathcal{G}_{AAS} -
\left\langle \bar{q}q\right\rangle \frac{\left(p^{2}+q^{2}-r^{2}\right)}
{p^{2}\,q^{2}} \,,\nonumber \\
\vbox{\vskip 6mm}
q_{\mu}\,p_{\nu}\,\Pi_{AAS}^{\mu\nu} &\,=\,&
-\,\frac{\lambda}{4}\,\mathcal{F}_{AAS} + \frac{\lambda}{8}\left(p^{2}+
q^{2}-r^{2}\right)\mathcal{G}_{AAS} + \left\langle \bar{q}q\right\rangle
\frac{\left(p^{2}+q^{2}-r^{2}\right)^{2}}{2\,p^{2}\,q^{2}} \,, \nonumber \\
\vbox{\vskip 6mm}
g_{\mu\nu}\,\Pi_{V\!AP}^{\mu\nu} &\,=\,&
\frac{3}{2}\left(p^{2}+q^{2}-r^{2}\right)\mathcal{F}_{V\!AP} -
\Big(\frac{\lambda}{4} + 3\,p^{2} q^{2}\Big)\mathcal{G}_{V\!AP} +
\left\langle \bar{q}q\right\rangle \frac{\left(p^{2}+5\,q^{2}-r^{2}\right)}
{q^{2}\,r^{2}} \,, \nonumber \\
\vbox{\vskip 6mm}
q_{\mu}\,p_{\nu}\,\Pi_{V\!AP}^{\mu\nu} &\,=\,&
-\,\frac{\lambda}{4}\,\mathcal{F}_{V\!AP} + \frac{\lambda}{8}\left(p^{2}+
q^{2}-r^{2}\right)\mathcal{G}_{V\!AP} - \left\langle \bar{q}q\right\rangle
\frac{\left(p^{2}-r^{2}\right)^{2}-q^{4}}{2\,q^{2}\,r^{2}} \,.\label{eq:traces}
\end{eqnarray}
Once these contractions are known, we are in a position to determine the scalar
functions $\mathcal{F}$ and $\mathcal{G}$ by inverting Eqs.~(\ref{eq:traces}).
The contractions have the same symmetry properties under exchange of momenta
as $\mathcal{F}$ and $\mathcal{G}$, due to Bose symmetry. Eqs.~(\ref{eq:traces})
admit an expansion in $\alpha_s$, and the terms proportional to
$\langle \bar{q}q\rangle$ contribute only at zeroth order in $\alpha_s$. Thus,
for the radiative corrections we are concerned about, the third and fifth lines
reduce to the first, while the fourth and sixth reduce to the second. The same
discussion concerning R$\chi$T and the OPE as in the odd--intrinsic--parity
sector applies also here, and again Wilson coefficients turn out to be scale
independent. We define those Wilson coefficients as
\begin{eqnarray}
g_{\mu\nu}\,\Pi_{VV\!S(AAS)[V\!AP]}^{\mu\nu}(\mu) &\,=\,&
C_{\langle\bar{q}q\rangle}^{g\, VV\!S(AAS)[V\!AP]}(p^{2},q^{2},r^{2})
\langle\bar{q}q\rangle(\mu) \,, \nonumber \\
\vbox{\vskip 6mm}
{q_{\mu}\,p_{\nu}\,\Pi}_{VV\!S(AAS)[V\!AP]}^{\mu\nu}(\mu) &\,=\,&
C_{\langle\bar{q}q\rangle}^{qp\, VV\!S(AAS)[V\!AP]}(p^{2},q^{2},r^{2})
\langle\bar{q}q\rangle(\mu) \,.
\end{eqnarray}

The leading--order results for these coefficients are found to be:
\begin{eqnarray}
C_{\langle\bar{q}q\rangle}^{g\,VV\!S} &\,=\,& \frac{1}{p^{2}\,q^{2}\,r^{2}}
\left[\left(p^{2}-q^{2}\right)^{2}+\left(p^{2}+q^{2}\right)r^2-2\,r^4\right] ,
\quad \hspace{1.8mm}
C_{\langle\bar{q}q\rangle}^{qp\,VV\!S} \,=\, -\,\frac{1}{2\, p^{2}\,q^{2}}\,
\lambda \,, \nonumber \\
\vbox{\vskip 6mm}
C_{\langle\bar{q}q\rangle}^{g\,AAS} &\,=\,& \frac{1}{p^{2}\,q^{2}\,r^{2}} 
\left[\left(p^{2}-q^{2}\right)^{2}-3\left(p^{2}+q^{2}\right)r^2+2\,r^4\right] ,
\quad\!
C_{\langle\bar{q}q\rangle}^{qp\,AAS} \,=\, \frac{1}{2\, p^{2}\,q^{2}}
\left(\lambda+4\, p^{2}q^{2}\right) , \nonumber \\
\vbox{\vskip 6mm}
C_{\langle\bar{q}q\rangle}^{g\,V\!AP} &\,=\,& -\,\frac{1}{p^{2}\,q^{2}\,r^{2}}
\left(p^{2}-q^{2}-2\,r^{2}\right)\left(p^{2}+q^{2}-\,r^{2}\right) , \nonumber\\
\vbox{\vskip 6mm}
C_{\langle\bar{q}q\rangle}^{qp\, V\!AP} &\,=\,& -\,\frac{1}{2\, p^{2}\,q^{2}\,
r^{2}}\left[r^{2}\lambda+2\, p^{2}\, q^{2}\left(p^{2}-q^{2}+r^{2}\right)
\right] .
\end{eqnarray}
Like before, the complete results of our computation at order $\alpha_s$ can
be found in Appendix~B.

\section{Conclusions\label{sec:Conclusions}}

In this work we have made a step forward in our understanding of quark--hadron
duality, by computing order $\alpha_{s}$ corrections to the leading quark
condensate contribution for 2-- and 3--point Green functions which are order
parameters of the chiral symmetry breaking. We have concentrated on these
order--parameter GFs, since in this case the matching between the OPE result
and the corresponding GFs obtained in a low--energy effective theory like
$\chi$PT or R$\chi$T is more transparent.

The matching further simplifies in the large--$N_{C}$ limit of QCD, because
then, the hadronic spectrum only consists of an infinite set of zero--width
resonances. Still, generally, at the next--to--leading order, the appearance
of logarithms in the kinematical variables requires the inclusion of the full
infinite number of those resonances in order to be able to reproduce the
logarithms present on the OPE side.

For the 2--point $\langle VT\rangle$ GF, we have discussed the matching in
more detail. The inclusion of $\alpha_{s}$ corrections allows to trace the
dependence of both, the OPE and the hadronic side, on the short--distance
renormalisation scale $\mu$. These $\mu$ dependencies show up in the quark
condensate as well as possible anomalous dimensions of the initial currents
on the OPE side, while in the low--energy theory, besides the condensate, the
LECs of the tensor sources explicitly depend on $\mu$ \cite{Cata-Mateu-1}.
To be able to do some numerics, finally we have assumed the
minimal--hadronic--ansatz as a crude approximation, and have matched the
$\langle VT\rangle$ GF with a single multiplet of vector--meson resonances.
This allowed to estimate the vector--meson tensor coupling $f_V^T$.

The main and most involved aim of our work was the calculation of the 
$\alpha_{s}$ corrections for the leading OPE term of 3--point order--parameter
GFs. Details of the computation for the different types of 3--point GFs to
be considered have been discussed in Section~4, while the rather lengthy
final results have been relegated to Appendix~B.

Additional complications for the matching in the case of 3--point GFs arise
from two facts: firstly, due to the presence of three independent kinematical
variables, no choice of the renormalisation scale $\mu$ allows to resum all
of the logarithms which are present in our results. Secondly, in addition in
the next--to--leading order result also dilogarithms arise and it is unclear
how to match them even with an infinite set of resonances on the hadronic side.
Therefore, we postpone a detailed discussion of the matching in the case of
3--point GFs to forthcoming work.

\bigskip
\acknowledgments
We would would like to thank Santi~Peris and Toni~Pich for useful discussions.
This work has been supported in part by the EU Contract No. MRTN-CT-2006-035482
(FLAVIAnet) (MJ, VM), by CICYT-FEDER-FPA2005-02211 (MJ), by MEC (Spain)
grant No. FPA2007-60323 (VM) and by Generalitat Valenciana grant No.
GVACOMP2007-156 (VM).

\section*{Appendix A: non--renormalisation of the $\mathbf{\langle AP\rangle}$
GF}

\addcontentsline{toc}{section}{Appendix A: non--renormalisation of the AP GF}\newcounter{alpha1} \renewcommand{\thesection}{\Alph{alpha1}} \renewcommand{\theequation}{\Alph{alpha1}.\arabic{equation}} \renewcommand{\thetable}{\Alph{alpha1}.\arabic{table}} \setcounter{alpha1}{1} \setcounter{equation}{0} \setcounter{table}{0}
In this appendix we demonstrate that in the chiral limit Ward identities
completely fix the $\langle AP\rangle$ GF, to be fully saturated by single
pion exchange. Lorentz and SU$(3)$ invariance imply
\begin{equation}
\left(\Pi_{AP}\right)_{\mu}^{ab}(p)\,=\, i\,\delta^{ab}\,\Pi_{AP}(p^{2})\, p_{\mu}\,.\label{eq:Loretz-AP}
\end{equation}
From the definition of time--ordered product and the conservation of the
axial--vector current it follows that
\begin{equation}
\partial^{\mu}\, T\{ A_{\mu}^{a}(x)\, P^{b}(0)\}\,=\,\delta(x^{0})\,[A_{0}^{a}(x),\, P^{b}(0)]\,.\label{eq:derivative}
\end{equation}
On the other hand Fourier transformation relates derivatives with external
momenta
\begin{equation}
\int\mathrm{d}^{4}x\, e^{x\cdot p}\left(\partial^{\mu}+i\, p^{\mu}\right)\left\langle \,0\left|\,\mathrm{T}\{ A_{\mu}^{a}(x)\, P^{b}(0)\}\right|0\,\right\rangle \,=\,0\,.
\end{equation}
The right hand side of Eq.~(\ref{eq:derivative}) can be easily computed
since the Dirac delta function forces the time dependence on both fields
to be the same:
\begin{equation}
\delta(x^{0}-y^{0})\left[A_{0}^{a}(x),P^{b}(y)\right]\,=\,-\, i\left\{ \!\dfrac{2}{n_{f}}\,\delta^{ab}\, S(x)
\,+\, d^{abc}\, S^{c}(x)\!
\right\} \,\delta^{(4)}(x-y)\,,\label{eq:conmutador}
\end{equation}
where $S(x)=\sum_{n_{f}}\bar{q}_{f}q_{f}$ is the flavour singlet
scalar current. Then from Eqs. (\ref{eq:Loretz-AP}) to (\ref{eq:APsat})
it follows that
\begin{equation}
p^{\mu}\left(\Pi_{AP}\right)_{\mu}^{ab}(p)\,=\, i\,\delta^{ab}\, p^{2}\,\Pi_{AP}\,=\, 
\dfrac{2\,i}{n_{f}}\,\delta^{ab}\left\langle \,0\left|\,S(0)\,\right|0\,\right\rangle 
\, =\,2\, i\,\delta^{ab}\left\langle \bar{q}q\right\rangle ,\label{eq:APsat}
\end{equation}
where we have used the fact that vacuum is flavour blind. We can identify $\Pi_{AP}=2\left\langle \bar{q}q\right\rangle /p^{2}$
and then demonstrate Eq.~(\ref{eq:pion-satur}).

\section*{Appendix B: explicit expression for the Wilson coefficients}

\addcontentsline{toc}{section}{Appendix B: explicit expression for the Wilson coefficients}\newcounter{alpha} \renewcommand{\thesection}{\Alph{alpha}} \renewcommand{\theequation}{\Alph{alpha}.\arabic{equation}} \renewcommand{\thetable}{\Alph{alpha}
} \setcounter{alpha}{2} \setcounter{equation}{0} \setcounter{table}{0}

\noindent In this appendix we present the complete expression for the
$\alpha_{s}$ corrections to the quark condensate Wilson coefficients. We
have split our results as shown in Eq.~(\ref{eq:3-P-C-gen}), in terms of
coefficients multiplying logarithms, dilogarithms and polynomial terms.
The results can be found in Table~\ref{WC-3p}.


\renewcommand{\arraystretch}{1.7}\setlength{\LTcapwidth}{\textwidth}

%

\tabletail{\hline}\tablehead{\hline}
\begin{longtable}[c]{|c|c|}
\hline
\endhead
\hline
\caption[$\alpha_{s}$ corrections to the Wilson coefficients for three--point GFs.]{\rule{0cm}{2em}$\alpha_{s}$ corrections to the Wilson coefficients for three--point GFs.}
\endfoot
\hline
\caption[$\alpha_{s}$ corrections to the Wilson coefficients for three--point GFs.]{\label{WC-3p}
\rule{0cm}{2em}$\alpha_{s}$ corrections to the Wilson coefficients for three--point GFs.}
\endlastfoot
\multicolumn{2}{|c|}{$\langle SSS\rangle$}\tabularnewline
\hline
$L_{p}$&
$\frac{4}{ p^{2}q^{2}r^{2}}\left[\,4\,p^{4}+q^{4}+r^{4}-6\,q^{2}\,r^{2}-3\,p^{2}\left(q^{2}+r^{2}\right)\right]$\tabularnewline
$L_{q}$&
$\frac{4}{ p^{2}q^{2}r^{2}}\left[\,4\,q^{4}+p^{4}+r^{4}-6\,p^{2}\,r^{2}-3\,q^{2}\left(p^{2}+r^{2}\right)\right]$\tabularnewline
$L_{r}$&
$\frac{4}{ p^{2}q^{2}r^{2}}\left[\,4\,r^{4}+q^{4}+p^{4}-6\,q^{2}\,p^{2}-3\,r^{2}\left(q^{2}+p^{2}\right)\right]$\tabularnewline
$L_{d}$&
$\frac{8}{ p^{2}q^{2}r^{2}}\left[\,p^{6}+q^{6}+r^{6}-2\,p^{4}\left(q^{2}+r^{2}\right)-2\,q^{4}\left(p^{2}+r^{2}\right)-2\,r^{4}\left(p^{2}+q^{2}\right)\right]$\tabularnewline
$L_{c}$&
$\frac{8}{ p^{2}q^{2}r^{2}}\left(-\,5\,p^{4}-5\,q^{4}-5\,r^{4}+14\,p^{2}\,r^{2}+14\,q^{2}\,p^{2}+14\,q^{2}\,r^{2}\right)$
\tabularnewline
\hline
\multicolumn{2}{|c|}{$\langle SPP\rangle$}\tabularnewline
\hline
$L_{p}$&
$\frac{4}{ p^{2}q^{2}r^{2}}\left[\,r^{4}+q^4-4\,p^4-3\,p^2\,(r^2+q^2)-6\,r^2\,q^2\,\right]$\tabularnewline
$L_{q}$&
$\frac{4}{ p^{2}q^{2}r^{2}}\left[\,4\,q^{4}+r^{4}-p^{4}+3\,q^{2}\left(p^{2}-r^{2}\right)\right]$\tabularnewline
$L_{r}$&
$\frac{4}{ p^{2}q^{2}r^{2}}\left[\,4\,r^{4}+q^{4}-p^{4}+3\,r^{2}\left(p^{2}-q^{2}\right)\right]$\tabularnewline
$L_{d}$&
$\frac{8}{ p^{2}q^{2}r^{2}}\left[\,r^{6}+q^{6}-p^{6}-2\,p^{2}\left(r^{4}+q^{4}\right)+2\left(p^{4}-q^{2}\,r^{2}\right)\left(r^{2}+q^{2}\right)\right]$\tabularnewline
$L_{c}$&
$\frac{8}{p^{2}q^{2}r^{2}}\left(-\,5\,r^{4}-5\,q^{4}+5\,p^{4}+14\,r^{2}\,q^{2}\right)$\tabularnewline
\hline
\multicolumn{2}{|c|}{$\langle VV\!P\rangle$}\tabularnewline
\hline
$L_{p}$&
$\frac{-\,2}{\lambda\, p^{2}q^{2}r^{2}}\left[\,p^{6}+r^{2}\,p^{4}+p^{2}\,q^{2}\left(5\,r^{2}+q^{2}\right)-2\left(r^{2}+q^{2}\right)\left(r^{2}-q^{2}\right)^{2}\right]$\tabularnewline
$L_{q}$&
$\frac{-\,2}{\lambda\, p^{2}q^{2}r^{2}}\left[\,q^{6}+r^{2}\,q^{4}+p^{2}\,q^{2}\left(5\,r^{2}+p^{2}\right)-2\left(r^{2}+p^{2}\right)\left(r^{2}-p^{2}\right)^{2}\right]$\tabularnewline
$L_{r}$&
$\frac{-\,2}{\lambda\,p^{2}q^{2}r^{2}}\left[\left(p^{2}+q^{2}\right)\left(p^{2}-q^{2}\right)^{2}+4\,r^{6}
-\,r^2\left(3\,p^{2}+q^{2}\right)\left(3\,q^{2}+p^{2}\right)-2\,r^4\left(q^{2}+p^{2}\right)\right]$\tabularnewline
$L_{d}$&
$\frac{-\,4}{\lambda\, p^{2}q^{2}r^{2}}\left[\,2\left(p^{4}-3\,p^{2}q^{2}+q^{4}\right)r^{4}-2\,r^2\left(p^{2}+q^{2}\right)\left(p^{4}+q^{4}\right)\right.$\tabularnewline
&
$\,\,\,\,\,\,\,\,\,\,\,\,\,\,\left.-\,2\,r^{6}\left(p^{2}+q^{2}\right)+r^{8}+\left(p^{2}-q^{2}\right)^{2}\left(p^{4}+q^{4}-q^{2}\,p^{2}\right)\right]$\tabularnewline
$L_{c}$&
$\frac{2}{ p^{2}q^{2}r^{2}}\left(p^2+q^2+4\,r^2\right)$\tabularnewline
\hline 
\multicolumn{2}{|c|}{$\langle AAP\rangle$}\tabularnewline
\hline
$L_{p}$&
$\frac{-\,2}{\lambda\, p^{2}q^{2}r^{2}}\left[\,p^{6}-3\,r^{2}\,p^{4}+p^{2}\,q^{2}\left(5\,r^{2}+q^{2}\right)+2\left(r^{2}-q^{2}\right)^{3}\right]$\tabularnewline
$L_{q}$&
$\frac{-\,2}{\lambda\, p^{2}q^{2}r^{2}}\left[\,q^{6}-3\,r^{2}\,q^{4}+p^{2}\,q^{2}\left(5\,r^{2}+p^{2}\right)+2\left(r^{2}-p^{2}\right)^{3}\right]$\tabularnewline
$L_{r}$&
$\frac{-\,2}{\lambda\, p^{2}q^{2}r^{2}}\left[-\,4\,r^{6}+6\,r^{4}\left(p^{2}+q^{2}\right)-r^2\left(q^{2}+3\,p^{2}\right)\left(p^{2}+3\,q^{2}\right)+\left(p^{2}-q^{2}\right)^{2}\left(p^{2}+q^{2}\right)\right]$\tabularnewline
$L_{d}$&
$\frac{-\,4}{\lambda\, p^{2}q^{2}r^{2}}\left[\,2\,r^{6}\left(p^{2}+q^{2}\right)+6\,p^{2}\,q^{2}\,r^{4}-2\,r^{2}\left(p^{2}+q^{2}\right)\left(p^{4}+q^{4}\right)\right.$\tabularnewline
&
$\left.-\,r^{8}+\left(p^{2}-q^{2}\right)^{2}\left(p^{4}+q^{4}-q^{2}\,p^{2}\right)\right]\,\,\,\,\,\,\,\,\,\,\,\,\,$\tabularnewline
$L_{c}$&
$\frac{2}{ p^{2}q^{2}r^{2}}\left(p^2+q^2-4\,r^2\right)$\tabularnewline
\hline 
\multicolumn{2}{|c|}{$\langle VA\,S\rangle$}\tabularnewline
\hline
$L_{p}$&
$\frac{-\,2}{\lambda\, p^{2}q^{2}r^{2}}\left[\,p^{6}+3\,p^{4}\left(2\,q^{2}-r^{2}\right)-p^{2}q^{2}\left(5\,r^{2}+q^{2}\right)+2\left(r^{2}+q^{2}\right)\left(r^{2}-q^{2}\right)^{2}\right]$\tabularnewline
$L_{q}$&
$\frac{-\,2}{\lambda\, p^{2}q^{2}r^{2}}\left[-\,2\,p^{6}+\left(q^{2}+6\,r^{2}\right)p^{4}-\left(6\,q^{4}+6\,r^{4}-5\,q^{2}\,r^{2}\right)p^{2}-q^{6}+2\,r^{6}-q^{4}\,r^{2}\,\right]$\tabularnewline
$L_{r}$&
$\frac{-\,2}{\lambda\, p^{2}q^{2}r^{2}}\left[\,p^{6}+7q^{2}\,p^{2}(q^{2}-p^{2})-q^{6}-4\,r^{6}+2\left(3\,p^{2}+q^{2}\right)r^{4}-3\left(p^{4}-q^{4}\right)r^{2}\,\right]$\tabularnewline
$L_{d}$&
$\frac{-\,4}{\lambda\, p^{2}q^{2}r^{2}}\left[\left(3\,q^{6}+2\,r^{2}\,q^{4}+6\,r^{4}\,q^{2}+2\,r^{6}\right)p^{2}-\left(r^{2}-q^{2}\right)^{2}\left(r^{4}+q^{4}\right)\right.$\tabularnewline
&
$+\,\left.p^{8}-p^{6}\left(3\,q^{2}+2\,r^{2}\right)-2\,r^{2}\,q^{2}\,p^{4}\,\right]\,\,\,\,\,\,\,\,\,\,\,\,\,\,\,\,\,\,\,\,\,\,\,\,\,\,\,\,\,\,\,\,\,\,$\tabularnewline
$L_{c}$&
$\frac{2}{ p^{2}q^{2}r^{2}}\left(p^2-q^2-4\,r^2\right)$\tabularnewline
\hline
\multicolumn{2}{|c|}{$g_{\mu\nu}\,\langle V^{\mu}V^{\nu}S\rangle$}\tabularnewline
\hline 
$L_{p}$&
$\frac{2}{ p^{2}q^{2}r^{2}}\left[\,2\left(q^{4}-r^{4}\right)-p^{4}+p^{2}\left(r^{2}-3\,q^{2}\right)\right]$\tabularnewline
$L_{q}$&
$\frac{2}{ p^{2}q^{2}r^{2}}\left[\,2\left(p^{4}-r^{4}\right)-q^{4}+q^{2}\left(r^{2}-3\,p^{2}\right)\right]$\tabularnewline
$L_{r}$&
$\frac{2}{ p^{2}q^{2}r^{2}}\left[\,6\,p^{2}\,q^{2}+4\,r^{4}-p^{4}-q^{4}-r^{2}\left(p^{2}+q^{2}\right)\right]$\tabularnewline
$L_{d}$&
$\frac{4}{ p^{2}q^{2}r^{2}}\left[\,2\,r^{6}-\left(r^{4}+p^{4}+q^{4}-3\,p^{2}\,q^{2}\right)\left(p^{2}+q^{2}\right)\right]$\tabularnewline
$L_{c}$&
$\frac{4}{p^{2}q^{2}r^{2}}\left[\,3\,r^2(p^2+q^2-r^2)-4\,p^2\,q^2 \right]$\tabularnewline
\hline
\multicolumn{2}{|c|}{$q_{\mu}\,p_{\nu}\,\langle V^{\mu}V^{\nu}S\rangle$}\tabularnewline
\hline
$L_{p}$&
$\frac{1}{ q^{2}r^{2}}\left[\,q^{2}\left(2\,q^{2}-2\,p^{2}+r^{2}\right)-3\,r^{2}\left(p^{2}-r^{2}\right)\right]$\tabularnewline
$L_{q}$&
$\frac{1}{ p^{2}r^{2}}\left[\,p^{2}\left(2\,p^{2}-2\,q^{2}+r^{2}\right)-3\,r^{2}\left(q^{2}-r^{2}\right)\right]$\tabularnewline
$L_{r}$&
$\frac{1}{p^{2}q^{2}}\left[3\,(p^4+q^4)-3\,r^2\,(p^2+q^2)-2\,p^2\,q^2\right]$\tabularnewline
$L_{d}$&
$\frac{2}{p^{2}q^{2}}\left[\,p^{6}-p^4\,r^2+(q^2-r^2)^2(q^2+r^2)-p^2\,r^2(4\,q^2+r^2)\right]$\tabularnewline
$L_{c}$&
$\frac{1}{q^{2}r^{2}p^{2}}\left(p^2+q^2-2\,r^2\right)\lambda$\tabularnewline
\hline 
\multicolumn{2}{|c|}{$g_{\mu\nu}\,\langle A^{\mu}A^{\nu}S\rangle$}\tabularnewline
\hline
$L_{p}$&
$\frac{2}{ p^{2}q^{2}r^{2}}\left[\,2\left(q^{2}-r^{2}\right)^{2}-p^{4}+p^{2}\left(5\,r^{2}-3\,q^{2}\right)\right]$\tabularnewline
$L_{q}$&
$\frac{2}{ p^{2}q^{2}r^{2}}\left[\,2\left(p^{2}-r^{2}\right)^{2}-q^{4}+q^{2}\left(5\,r^{2}-3\,p^{2}\right)\right]$\tabularnewline
$L_{r}$&
$\frac{2}{ p^{2}q^{2}r^{2}}\left[\,6\,p^{2}\,q^{2}-4\,r^{4}-p^{4}-q^{4}-r^{2}\left(p^{2}+q^{2}\right)\right]$\tabularnewline
$L_{d}$&
$-\frac{4}{ p^{2}q^{2}r^{2}}\left[\,2\,r^{6}+\left(-\,3\,r^{4}+p^{4}+q^{4}-3\,p^{2}\,q^{2}\right)\left(p^{2}+q^{2}\right)\right]$\tabularnewline
$L_{c}$&
$\frac{4}{p^{2}q^{2}r^{2}}\left[\,3\,r^{4}-4\,p^{2}\,q^{2}-3\,r^{2}\left(q^{2}+p^{2}\right)\right]$\tabularnewline
\hline 
\multicolumn{2}{|c|}{$q_{\mu}\,p_{\nu}\,\langle A^{\mu}A^{\nu}S\rangle$}\tabularnewline
\hline
$L_{p}$&
$\frac{1}{ q^{2}r^{2}}\left[\,r^{2}\left(q^{2}-3\,p^{2}+3\,r^{2}\right)+2\,q^{2}\left(q^{2}-p^{2}\right)\right]$\tabularnewline
$L_{q}$&
$\frac{1}{ p^{2}r^{2}}\left[\,r^{2}\left(p^{2}-3\,q^{2}+3\,r^{2}\right)-2\,p^{2}\left(q^{2}-p^{2}\right)\right]$\tabularnewline
$L_{r}$&
$\frac{1}{ p^{2}q^{2}}\left[-\,2\,p^2\,q^2+3\,(p^4+q^4)-3\,r^2(p^2+q^2)\right]$\tabularnewline
$L_{d}$&
$\frac{2}{ p^{2}q^{2}}\left[\,p^{6}-3\,p^4\,r^2+3\,p^2\,r^4+(q^2-r^2)^3\right]$\tabularnewline
$L_{c}$&
$\frac{1}{ p^{2}q^{2}r^{2}}\left(p^2+q^2+2\,r^2\right)\lambda$\tabularnewline
\hline
\multicolumn{2}{|c|}{$g_{\mu\nu}\,\langle V^{\mu}A^{\nu}P\rangle$}\tabularnewline
\hline
$L_{p}$&
$\frac{2}{ p^{2}q^{2}r^{2}}\left[\,2\left(q^{4}-r^{4}\right)+p^{4}-p^{2}\left(5\,r^{2}+q^{2}\right)\right]$\tabularnewline
$L_{q}$&
$\frac{2}{ p^{2}q^{2}r^{2}}\left[\,p^{2}\left(q^{2}+4\,r^{2}\right)-2\,p^{4}-q^{4}-2\,r^{4}+q^{2}\,r^{2}\right]$\tabularnewline
$L_{r}$&
$\frac{2}{ p^{2}q^{2}r^{2}}\left[\,p^{4}-q^{4}+4\,r^{4}+r^{2}\left(p^{2}-q^{2}\right)\right]$\tabularnewline
$L_{d}$&
$\frac{4}{p^{2}q^{2}r^{2}}\left[\,p^{6}-2\,q^{2}\,p^{4}+2\,p^{2}\,q^{4}-q^{6}+2\,r^{6}-r^{4}\left(3\,p^{2}+q^{2}\right)\right]$\tabularnewline
$L_{c}$&
$\frac{12}{p^{2}q^{2}}\left(q^{2}+p^{2}-r^{2}\right)$\tabularnewline
\hline 
\multicolumn{2}{|c|}{$q_{\mu}\,p_{\nu}\,\langle V^{\mu}A^{\nu}P\rangle$}\tabularnewline
\hline
$L_{p}$&
$\frac{3}{q^{2}}\left(p^{2}+3\,q^{2}-r^{2}\right)$\tabularnewline
$L_{q}$&
$-\,\frac{3}{ p^{2}}\left(3\,p^{2}+q^{2}-r^{2}\right)$\tabularnewline
$L_{r}$&
$\frac{3}{ p^{2}q^{2}}\left(q^{2}-p^{2}\right)\left(q^{2}+p^{2}-r^{2}\right)$\tabularnewline
$L_{d}$&
$-\frac{2}{ p^{2}q^{2}}\left[\,p^6-3\, r^2\, p^4+ p^2\,r^2 \left(2\, q^2+3\, r^2\right)-\left(q^2-r^2\right)^2
   \left(q^2+r^2\right)\right]$\tabularnewline
$L_{c}$&
$-\,\frac{1}{ p^{2}q^{2}r^{2}}\left(p^2-q^2+2\,r^2\right)\lambda$\tabularnewline

\end{longtable}

\bigskip

\begin{thebibliography}{10}

\bibitem{Weinberg}
S.~Weinberg, {\it {Phenomenological Lagrangians}},  {\em Physica} {\bf A96}
  (1979) 327.

\bibitem{Gasser1}
J.~Gasser and H.~Leutwyler, {\it {Chiral Perturbation Theory to One Loop}},
  {\em Ann. Phys.} {\bf 158} (1984) 142.

\bibitem{Gasser2}
J.~Gasser and H.~Leutwyler, {\it {Chiral Perturbation Theory: Expansions in the
  Mass of the Strange Quark}},  {\em Nucl. Phys.} {\bf B250} (1985) 465.

\bibitem{OPEwilson}
K.~G. Wilson, {\it {Non--lagrangian Models of Current Algebra}},  {\em Phys.
  Rev.} {\bf 179} (1969) 1499--1512.

\bibitem{OPE-ITEP}
M.~A. Shifman, A.~I. Vainshtein, and V.~I. Zakharov, {\it {QCD and Resonance
  Physics: Sum Rules}},  {\em Nucl. Phys.} {\bf B147} (1979) 385--447.

\bibitem{SVZ2}
M.~A. Shifman, A.~I. Vainshtein, and V.~I. Zakharov, {\it {QCD and Resonance
  Physics: Applications}},  {\em Nucl. Phys.} {\bf B147} (1979) 448--518.

\bibitem{largeN1}
G.~'t~Hooft, {\it {A Planar Diagram Theory for Strong Interactions}},  {\em
  Nucl. Phys.} {\bf B72} (1974) 461.

\bibitem{largeN2}
G.~'t~Hooft, {\it {A Two-Dimensional Model for Mesons}},  {\em Nucl. Phys.}
  {\bf B75} (1974) 461.

\bibitem{largeN3}
E.~Witten, {\it {Baryons in the $1/N$ Expansion}},  {\em Nucl. Phys.} {\bf
  B160} (1979) 57.

\bibitem{RChT1}
G.~Ecker, J.~Gasser, A.~Pich, and E.~de~Rafael, {\it {The Role of Resonances in
  Chiral Perturbation Theory}},  {\em Nucl. Phys.} {\bf B321} (1989) 311.

\bibitem{RChT2}
G.~Ecker, J.~Gasser, H.~Leutwyler, A.~Pich, and E.~de~Rafael, {\it {Chiral
  Lagrangians for Massive Spin 1 Fields}},  {\em Phys. Lett.} {\bf B223} (1989)
  425.

\bibitem{Moussallam:1997xx}
B.~Moussallam, {\it {A sum rule approach to the violation of Dashen's
  theorem}},  {\em Nucl. Phys.} {\bf B504} (1997) 381--414,
  [\href{http://xxx.lanl.gov/abs/hep-ph/9701400}{{\tt hep-ph/9701400}}].

\bibitem{Peris:1998nj}
S.~Peris, M.~Perrottet, and E.~de~Rafael, {\it {Matching long and short
  distances in large-$N_C$ {QCD}}},  {\em JHEP} {\bf 05} (1998) 011,
  [\href{http://xxx.lanl.gov/abs/hep-ph/9805442}{{\tt hep-ph/9805442}}].

\bibitem{MP07}
P.~Masjuan and S.~Peris, {\it {A Rational Approach to Resonance Saturation in
  large-$N_C$ QCD}},  {\em JHEP} {\bf 05} (2007) 040,
  [\href{http://xxx.lanl.gov/abs/arXiv:0704.1247 [hep-ph]}{{\tt arXiv:0704.1247
  [hep-ph]}}].

\bibitem{mondejar}
J.~Mondejar and A.~Pineda, {\it {Constraints on Regge models from perturbation
  theory}},  {\em JHEP} {\bf 10} (2007) 061,
  [\href{http://xxx.lanl.gov/abs/arXiv:0704.1417 [hep-ph]}{{\tt arXiv:0704.1417
  [hep-ph]}}].

\bibitem{Knecht:2001xc}
M.~Knecht and A.~Nyffeler, {\it {Resonance estimates of ${\cal O}(p^6)$
  low-energy constants and QCD short-distance constraints}},  {\em Eur. Phys.
  J.} {\bf C21} (2001) 659--678,
  [\href{http://xxx.lanl.gov/abs/hep-ph/0106034}{{\tt hep-ph/0106034}}].

\bibitem{RuizFemenia:2003hm}
P.~D. Ruiz-Femenia, A.~Pich, and J.~Portol\'es, {\it {Odd-intrinsic-parity
  processes within the resonance effective theory of QCD}},  {\em JHEP} {\bf
  07} (2003) 003, [\href{http://xxx.lanl.gov/abs/hep-ph/0306157}{{\tt
  hep-ph/0306157}}].

\bibitem{Cirigliano:2004ue}
V.~Cirigliano, G.~Ecker, M.~Eidem{\"u}ller, A.~Pich, and J.~Portol\'es, {\it
  {The $\langle V\!AP\rangle$ Green function in the resonance region}},  {\em
  Phys. Lett.} {\bf B596} (2004) 96--106,
  [\href{http://xxx.lanl.gov/abs/hep-ph/0404004}{{\tt hep-ph/0404004}}].

\bibitem{SPP}
V.~Cirigliano {\em et~al.}, {\it {The $\langle SPP\rangle$ Green function and
  SU(3) breaking in $K_{l3}$ decays}},  {\em JHEP} {\bf 04} (2005) 006,
  [\href{http://xxx.lanl.gov/abs/hep-ph/0503108}{{\tt hep-ph/0503108}}].

\bibitem{Cirigliano:2006hb}
V.~Cirigliano {\em et~al.}, {\it {Towards a consistent estimate of the chiral
  low-energy constants}},  {\em Nucl. Phys.} {\bf B753} (2006) 139--177,
  [\href{http://xxx.lanl.gov/abs/hep-ph/0603205}{{\tt hep-ph/0603205}}].

\bibitem{Mateu-Portoles}
V.~Mateu and J.~Portol\'es, {\it {Form Factors in radiative pion decay}},  {\em
  Eur. Phys. J.} {\bf C52} (2007) 325--338,
  [\href{http://xxx.lanl.gov/abs/arXiv:0706.1039 [hep-ph]}{{\tt arXiv:0706.1039
  [hep-ph]}}].

\bibitem{Nonqq1}
V.~P. Spiridonov and K.~G. Chetyrkin, {\it {Nonleading mass corrections and
  renormalization of the operators $m\bar\psi\psi$ and $G_{\mu\nu}^2$}},  {\em
  Sov. J. Nucl. Phys.} {\bf 47} (1988) 522--527.

\bibitem{Nonqq2}
M.~Jamin and M.~M{\"u}nz, {\it {Current correlators to all orders in the quark
  masses}},  {\em Z. Phys.} {\bf C60} (1993) 569--578,
  [\href{http://xxx.lanl.gov/abs/hep-ph/9208201}{{\tt hep-ph/9208201}}].

\bibitem{Nonqq3}
M.~Jamin, {\it {Flavour-symmetry breaking of the quark condensate and chiral
  corrections to the Gell-Mann-Oakes-Renner relation}},  {\em Phys. Lett.} {\bf
  B538} (2002) 71--76, [\href{http://xxx.lanl.gov/abs/hep-ph/0201174}{{\tt
  hep-ph/0201174}}].

\bibitem{BJW90}
A.~J. Buras, M.~Jamin, and P.~H. Weisz, {\it {Leading and Next-to-Leading QCD
  Corrections to $\varepsilon$-Parameter and $B_{0}$--$\bar B_{0}$ Mixing in
  the Presence of a Heavy Top Quark}},  {\em Nucl. Phys.} {\bf B347} (1990)
  491--536.

\bibitem{BG95}
D.~J. Broadhurst and A.~G. Grozin, {\it {Matching QCD and HQET heavy - light
  currents at two loops and beyond}},  {\em Phys. Rev.} {\bf D52} (1995)
  4082--4098, [\href{http://xxx.lanl.gov/abs/hep-ph/9410240}{{\tt
  hep-ph/9410240}}].

\bibitem{Gra00}
J.~A. Gracey, {\it {Three--loop $\overline{\rm MS}$ tensor current anomalous
  dimension in QCD}},  {\em Phys. Lett.} {\bf B488} (2000) 175--181,
  [\href{http://xxx.lanl.gov/abs/hep-ph/0007171}{{\tt hep-ph/0007171}}].

\bibitem{Cata:2008zc}
O.~Cata and V.~Mateu, {\it {Novel patterns for vector mesons from the
  large-$N_C$ limit}},  \href{http://xxx.lanl.gov/abs/arXiv:0801.4374
  [hep-ph]}{{\tt arXiv:0801.4374 [hep-ph]}}.

\bibitem{Cata-Mateu-1}
O.~Cata and V.~Mateu, {\it {Chiral Perturbation Theory with Tensor Sources}},
  {\em JHEP} {\bf 09} (2007) 078,
  [\href{http://xxx.lanl.gov/abs/arXiv:0705.2948 [hep-ph]}{{\tt arXiv:0705.2948
  [hep-ph]}}].

\bibitem{BB96}
P.~Ball and V.~M. Braun, {\it {The $\rho$ Meson Light-Cone Distribution
  Amplitudes of Leading Twist Revisited}},  {\em Phys. Rev.} {\bf D54} (1996)
  2182--2193, [\href{http://xxx.lanl.gov/abs/hep-ph/9602323}{{\tt
  hep-ph/9602323}}].

\bibitem{Braunetal03}
V.~M. Braun {\em et~al.}, {\it {A lattice calculation of vector meson couplings
  to the vector and tensor currents using chirally improved fermions}},  {\em
  Phys. Rev.} {\bf D68} (2003) 054501,
  [\href{http://xxx.lanl.gov/abs/hep-lat/0306006}{{\tt hep-lat/0306006}}].

\bibitem{Narison1}
S.~Narison and N.~Paver, {\it {On some Three Meson Vertex Sum Rules in Quantum
  Chromodynamics}},  {\em Z. Phys.} {\bf C22} (1984) 69.

\bibitem{Narison2}
S.~Narison, {\it {On the Two Photon Width of the $\delta(980)$}},  {\em Phys.
  Lett.} {\bf B175} (1986) 88.

\bibitem{tHV79}
G.~'t~Hooft and M.~J.~G. Veltman, {\it {Scalar One Loop Integrals}},  {\em
  Nucl. Phys.} {\bf B153} (1979) 365--401.

\bibitem{BGLP03}
J.~Bijnens, E.~G\'amiz, E.~Lipartia, and J.~Prades, {\it {QCD short-distance
  constraints and hadronic approximations}},  {\em JHEP} {\bf 04} (2003) 055,
  [\href{http://xxx.lanl.gov/abs/hep-ph/0304222}{{\tt hep-ph/0304222}}].

\end{thebibliography}

\providecommand{\href}[2]{#2}\begingroup\raggedright\endgroup

\end{document}